\documentstyle[eqsecnum,preprint,aps,tighten]{revtex}

\newtheorem{lemma}{Lemma}
\newtheorem{proposition}{Proposition}
\newtheorem{theorem}{Theorem}

\begin{document}

\draft

\preprint{UTPT-95-05}

\title{Regularity Theorems in the Nonsymmetric Gravitational Theory}

\author{J.W. Moffat}

\address{Department of Physics, University of Toronto,
Toronto, Ontario, Canada M5S 1A7}
\address{Department of Physics, Cave Hill Campus,
University of West Indies, Barbados}

\date{\today}

\maketitle

\begin{abstract}
Regularity theorems are presented for cosmology and gravitational collapse
in non-Riemannian gravitational theories. These theorems
establish conditions necessary to allow the existence of timelike and null
path complete spacetimes for matter that satisfies the positive
energy condition.  Non-Riemannian theories of gravity can have solutions
that have a non-singular beginning of the universe, and the gravitational
collapse of a star does not lead to a black hole event horizon and a
singularity as a final stage of collapse.
A perturbatively consistent version of nonsymmetric
gravitational theory is studied that, in the long-range approximation, has
a nonsingular static spherically
symmetric solution which is path complete, does not have black
hole event horizons and has finite curvature invariants.
The theory satisfies
the regularity theorems for cosmology and gravitational collapse.
The elimination of black holes resolves the information loss puzzle.
\end{abstract}

\pacs{}

\section{Introduction}
Einstein's gravitational theory (EGT)
\cite{bib:Einstein_1}
is an aesthetically pleasing
theory, which agrees with all observational tests to date. However, all these
tests are for weak gravitational fields; there is no convincing
evidence that the theory will be valid for strong gravitational fields.
The occurrence of singularities in the theory, when physical quantities such
as the density and pressure of matter become infinite, produces an
unphysical feature of the theory. The theorems of Hawking and Penrose
\cite{bib:Penrose,bib:Hawking_1,bib:Hawking_2,bib:Hawking_3,bib:Geroch,bib:Wald}
prove that for physical matter in EGT, these singularities cannot be
avoided.

A new version of the nonsymmetric gravitational theory (NGT) has recently
been proposed which is perturbatively consistent
\cite{bib:Moffat_1,bib:Moffat_2,bib:Legare}.
It is free of ghost poles and tachyons in the linear approximation and in
a linear expansion of $g_{[\mu\nu]}$ about a curved Einstein background.
The static spherically symmetric Wyman solution of the NGT vacuum field
equations
was found to be nonsingular with no event horizons and with
finite curvature invariants
\cite{bib:Cornish_1,bib:Cornish_2}.
These nonsingular results are
valid in the long-range approximation of the new version of NGT. There are
two basic parameters in the solution that enter as constants of integration:
the conserved gravitational charge $m$ and a dimensionless constant $s$.
The limit to EGT, as $s\rightarrow 0$, is not analytic at points $r < 2m$.
A final collapsed object (FCO), which is stable for an arbitrarily large mass,
replaces the black hole
\cite{bib:Cornish_1,bib:Cornish_2,bib:Cornish_3}.

The problem of black hole information loss, first considered in
detail by Hawking, has received wide attention recently
\cite{bib:Hawking_4}.
Since
spacetime is divided by an event horizon into two causally disconnected
manifolds, and since Hawking radiation is approximately thermal, all the
information
associated with a collapsing star which passes through the event horizon is
lost to an outside observer. After
the black hole has radiated away to nothing by the process of Hawking
radiation, all information about the star will have disappeared. Since
an initial
pure quantum state before collapse becomes a mixed state in the final stage
of collapse, the unitarity of the scattering S-matrix is violated.
A possible way out of this breakdown of the predictability of physics
is to modify EGT, so that the event horizon and the singularity
which occur as final features of gravitational collapse are removed.
This would solve the information loss puzzle in a simple way at the
classical level.

It would seem that we are faced with an almost insurmountable problem
when attempting to discover a classical modification of EGT, that removes
all its singular features.
The reason for this is that such a modification might
be expected to lead to a solution of the field equations, which can be
mapped into one that lies near the Schwarzschild solution of EGT,
and shares its
singular properties. Given that the extra degrees of freedom
associated with the modified gravitational theory are related to
certain fields and coupling constants, if the limit to
EGT is smooth and analytic in the coupling constants, then it would follow
that a set of initial data on a Cauchy surface can always be found that
leads, on solving the field equations of the theory, to a black hole
as the final stage of collapse. Therefore,
the modified theory of gravity should have a global exact solution for
collapse which is non-perturbative, and does not have a smooth
limit in the strong gravitational field regime at the event
horizon radius, $r=2m$, and at $r=0$. This would mean that there is no
smoooth limit to the Schwarzschild solution for $r\leq 2m$. Even for a
very large black hole
with weak curvature at $r=2m$, the modified theory must not display a
null surface (event horizon), so we must require that the solution
of the modified field equations be non-perturbative and non-analytic in the
new degrees of freedom at $r \leq 2m$. Of course, for large values
of $r$, corresponding to weak gravitational fields, the solution
should be smoothly analytic in the coupling constants in the EGT limit,
in order to retain the agreement of EGT with observations.
This is the case for the nonsingular Wyman solution of NGT
\cite{bib:Cornish_1,bib:Cornish_2,bib:Sokolov}.

We shall study the necessary conditions that must be
satisfied by NGT to guarantee the existence of regular solutions in
cosmology and gravitational collapse. Because NGT is based on
non-Riemannian
geometry, it is possible to satisfy a generalized positive energy condition
and permit the existence of timelike and null path complete
spacetimes. Thus, in a non-Riemannian geometry, the singularity theorems of
Hawking and Penrose can be circumvented for physical matter in strongly
causal spacetimes. In contrast to Einstein's theory of gravity, the
gravitational field can be repulsive for physical matter fields. NGT
is among a class of non-Riemannian theories that possess this feature.
The non-singular static Wyman solution of the NGT
field equations does not contain any trapped surfaces for $r < 2m$, and
it is shown to satisfy the regularity theorem for gravitational collapse.

Another class of theories that can circumvent the Hawking-Penrose theorems
for positive density and non-negative pressure is higher-derivative
theories based on Riemannian geometry
\cite{bib:Weyl,bib:Stelle,bib:Mannheim}.
These theories share
the feature with
NGT that gravity is not always attractive, so that the focussing effect
of geodesics typical of gravity theories based on Riemannian geometry
can be removed, thus allowing
singularity-free solutions for gravitational collapse. However, these theories
generally have ghost poles, tachyons or second-order poles in the linear
approximation when they are expanded about Minkowski spacetime.
Thus, these theories lead to unstable solutions and negative energy modes.

Another alternative theory of gravitation has been proposed by Yilmaz
\cite{bib:Yilmaz}.
This theory contains an energy-momentum tensor density defined in terms
of a scalar field $\phi$, which acts as a source for the gravitational field.
The static spherically symmetric solution is free of event horizons,
but the singularity at $r=0$ remains as an essential singularity in the
metric tensor. This is an undesirable feature of a gravitational theory,
because the absence of an event horizon in Yilmaz's theory produces a naked
singularity which destroys the Cauchy solution and any predictable power of
the theory. This feature is shared by Einstein's gravitational theory
for solutions which do not have black hole event horizons and by
any theory which does not remove the essential singularity at $r=0$. In NGT,
there is no singularity at $r=0$ and the removal of the black hole event
horizon does not destroy the predictability of the theory.

In Section 2, we present a perturbatively consistent NGT based on
non-Riemannian
geometry, while in Section 3 we give useful formulas connected with
affine connections and curvature tensors. We determine, in Section 4, the
properties of the geodesic and path equations of
motion in NGT, and discuss the geodesic and path deviation equations. The
static spherically symmetric solution of the NGT field equations
is treated in Section 5, while Section 6 is devoted to the
properties of geodesic and path congruences and the consequences of the
generalized Raychaudhuri equation.
In Sections 7 and 8, regularity theorems in cosmology and gravitational
collapse
are proved, and we present concluding remarks in Section 9.

\section{Nonsymmetric Gravitational Theory}

The non-Riemannian geometry is based on the nonsymmetric field structure with a
nonsymmetric $g_{\mu\nu}$
\cite{bib:Einstein_2,bib:Moffat_3}
\begin{equation}
g_{\mu\nu}=g_{(\mu\nu)}+g_{[\mu\nu]},
\end{equation}
where
\begin{equation}
g_{(\mu\nu)}=\frac{1}{2}(g_{\mu\nu}+g_{\nu\mu}),\quad g_{[\mu\nu]}=
\frac{1}{2}(g_{\mu\nu}-g_{\nu\mu}).
\end{equation}
Moreover, the affine connection is nonsymmetric:
\begin{equation}
\Gamma^\lambda_{\mu\nu}=\Gamma^\lambda_{(\mu\nu)}
+\Gamma^\lambda_{[\mu\nu]}.
\end{equation}
The contravariant tensor $g^{\mu\nu}$ is defined in terms of the equation:
\begin{equation}
g^{\mu\nu}g_{\sigma\nu}=g^{\nu\mu}g_{\nu\sigma}=\delta^\mu_\sigma.
\end{equation}

The Lagrangian density is given by
\begin{equation}
{\cal L}_{NGT}={\cal L}_R+{\cal L}_M,
\end{equation}
where
\begin{equation}
{\cal L}_R={\bf g}^{\mu\nu}R_{\mu\nu}(W)-2\Lambda\sqrt{-g}
-\frac{1}{4}\mu^2{\bf g}^{\mu\nu}g_{[\nu\mu]}-\frac{1}{6}{\bf g}^{\mu\nu}
W_\mu W_\nu,
\end{equation}
where $\Lambda$ is the cosmological constant and $\mu^2$ is the square
of a mass associated with $g_{[\mu\nu]}$.
Moreover, ${\cal L}_M$ is the matter Lagrangian density ($G=c=1$):
\begin{equation}
{\cal L}_M=-8\pi g^{\mu\nu}{\bf T}_{\mu\nu}.
\end{equation}
Here, ${\bf g}^{\mu\nu}=\sqrt{-g}g^{\mu\nu}$ and $R_{\mu\nu}(W)$ is the
NGT contracted curvature tensor:
\begin{equation}
R_{\mu\nu}(W)=W^\beta_{\mu\nu,\beta} - \frac{1}{2}
(W^\beta_{\mu\beta,\nu}+W^\beta_{\nu\beta,\mu}) -
W^\beta_{\alpha\nu}W^\alpha_{\mu\beta} +
W^\beta_{\alpha\beta}W^\alpha_{\mu\nu},
\end{equation}
defined in terms of the unconstrained nonsymmetric connection:
\begin{equation}
\label{eq:2.9}
W^\lambda_{\mu\nu}=\Gamma^\lambda_{\mu\nu}-\frac{2}{3}\delta^\lambda_\mu
W_\nu,
\end{equation}
where
\begin{equation}
W_\mu\equiv W^\lambda_{[\mu\lambda]}=
\frac{1}{2}(W^\lambda_{\mu\lambda}-W^\lambda_{\lambda\mu}).
\end{equation}
(\ref{eq:2.9}) leads to the result:
\begin{equation}
\label{eq:2.11}
\Gamma_\mu=\Gamma^\lambda_{[\mu\lambda]}=0.
\end{equation}

The NGT contracted curvature tensor can be written as
\begin{equation}
R_{\mu\nu}(W) = R_{\mu\nu}(\Gamma) + \frac{2}{3} W_{[\mu,\nu]},
\end{equation}
where $R_{\mu\nu}(\Gamma)$ is defined by
\begin{equation}
\label{eq:2.13}
R_{\mu\nu}(\Gamma ) = \Gamma^\beta_{\mu\nu,\beta} -\frac{1}{2}
\left(\Gamma^\beta_{(\mu\beta),\nu} + \Gamma^\beta_{(\nu\beta),\mu}\right) -
\Gamma^\beta_{\alpha\nu} \Gamma^\alpha_{\mu\beta} +
\Gamma^\beta_{(\alpha\beta)}\Gamma^\alpha_{\mu\nu}.
\end{equation}

The field equations in the presence of matter sources are given by:
\begin{equation}
\label{eq:2.14}
G_{\mu\nu} (W)+\Lambda g_{\mu\nu}+\frac{1}{4}\mu^2C_{\mu\nu}
-\frac{1}{6}(P_{\mu\nu}-\frac{1}{2}g_{\mu\nu}P) = 8\pi T_{\mu\nu},
\end{equation}
\begin{equation}
\label{eq:2.15}
{{\bf g}^{[\mu\nu]}}_{,\nu} = -\frac{1}{2}{\bf g}^{(\mu\beta)}W_\beta,
\end{equation}
\begin{eqnarray}
&{{\bf g}^{\mu\nu}}_{,\sigma}+{\bf g}^{\rho\nu}W^\mu_{\rho\sigma}
+{\bf g}^{\mu\rho}
W^\nu_{\sigma\rho}-{\bf g}^{\mu\nu}W^\rho_{\sigma\rho}
+\frac{2}{3}\delta^\nu_\sigma{\bf g}^{\mu\rho}W^\beta_{[\rho\beta]}
& \nonumber \\
\label{eq:2.16}
&\mbox{}+\frac{1}{6}({\bf g}^{(\mu\beta)}W_\beta\delta^\nu_\sigma
-{\bf g}^{(\nu\beta)}W_\beta\delta^\mu_\sigma)=0. &
\end{eqnarray}
Here, we have
\begin{equation}
G_{\mu\nu} = R_{\mu\nu} - {1\over 2} g_{\mu\nu} R,
\end{equation}
\begin{equation}
C_{\mu\nu}=g_{[\mu\nu]}+{1\over 2}g_{\mu\nu}g^{[\sigma\rho]}
g_{[\rho\sigma]}+g^{[\sigma\rho]}g_{\mu\sigma}g_{\rho\nu},
\end{equation}
\begin{equation}
P_{\mu\nu}=W_\mu W_\nu,
\end{equation}
and $P=g^{\mu\nu}P_{\mu\nu}=g^{(\mu\nu)}W_\mu W_\nu$.

With the help of (\ref{eq:2.9}) and the relation obtained from
(\ref{eq:2.15}):
\begin{equation}
\label{eq:2.20}
W_\mu=-\frac{2}{\sqrt{-g}}s_{\mu\rho}{{\bf g}^{[\rho\sigma]}}_{,\sigma},
\end{equation}
where $s_{\mu\rho}$ is defined by
\begin{equation}
s_{\mu\rho}g^{(\rho\nu)}=s_{\mu\rho}s^{\rho\nu}=\delta^\nu_\mu,
\end{equation}
we can recast (\ref{eq:2.16}) in the form:
\begin{equation}
\label{eq:2.22}
g_{\mu\nu,\sigma}-g_{\rho\nu}\Gamma^\rho_{\mu\sigma}
-g_{\mu\rho}\Gamma^\rho_{\sigma\nu}=Y^\rho\Lambda_{\mu\nu\sigma\rho},
\end{equation}
where
\begin{equation}
Y^\rho=\frac{1}{3\sqrt{-g}}{{\bf g}^{[\rho\sigma]}}_{,\sigma},
\end{equation}
and
\begin{equation}
\Lambda_{\mu\nu\sigma\rho}=g_{\mu\rho}g_{\sigma\nu}-g_{\mu\sigma}g_{\rho\nu}
-g_{\mu\nu}g_{[\sigma\rho]}.
\end{equation}
The field equations (\ref{eq:2.14}) can be written:
\begin{eqnarray}
&S_{\mu\nu}\equiv R_{\mu\nu}(\Gamma)-\frac{4}{3\sqrt{-g}}
(s_{[\mu\rho}{{\bf g}^{[\rho\sigma]}}_{,\sigma})_{,\nu]}
+\Lambda g_{\mu\nu} & \nonumber \\
\label{eq:2.25}
&\mbox{}+\frac{1}{4}\mu^2(C_{\mu\nu}-\frac{1}{2}g_{\mu\nu}C)
-\frac{1}{6}P_{\mu\nu}=8\pi(T_{\mu\nu}-\frac{1}{2}g_{\mu\nu}T), &
\end{eqnarray}
where $C=g^{\mu\nu}C_{\mu\nu}$ and we now have
\begin{equation}
P_{\mu\nu}=-\frac{4}{g}s_{\mu\rho}s_{\nu\tau}
{{\bf g}^{[\rho\sigma]}}_{,\sigma}{{\bf g}^{[\tau\alpha]}}_{,\alpha}.
\end{equation}
The generalized Bianchi identities:
\begin{equation}
[{\bf g}^{\alpha\nu}G_{\rho\nu}(\Gamma)+{\bf g}^{\nu\alpha}
G_{\nu\rho}(\Gamma)]_{,\alpha}+{g^{\mu\nu}}_{,\rho}{\bf G}_{\mu\nu}=0,
\end{equation}
give rise to the matter response equations:
\begin{equation}
g_{\mu\rho}{{\bf T}^{\mu\nu}}_{,\nu}+g_{\rho\mu}{{\bf T}^{\nu\mu}}_{,\nu}
+(g_{\mu\rho,\nu}+g_{\rho\nu,\mu}-g_{\mu\nu,\rho}){\bf T}^{\mu\nu}=0.
\end{equation}

The Christoffel symbols, in NGT, are defined by
\begin{equation}
\label{eq:2.29}
\left\{{\lambda\atop \mu\nu}\right\}=\frac{1}{2}s^{\lambda\rho}
\left(s_{\mu\rho,\nu}+s_{\rho\nu,\mu}-s_{\mu\nu,\rho}\right).
\end{equation}

\section{Affine Connections and Curvature Tensors}\par

We shall adopt the covariant derivative notation:
\begin{equation}
D_\mu v^\lambda={v^\lambda}_{,\mu}+v^\rho\Gamma^\lambda_{\rho\mu},\quad
D_\mu v_\lambda=v_{\lambda,\mu}-v_\rho\Gamma^\rho_{\lambda\mu},
\end{equation}
and
\begin{equation}
\nabla_\mu v^\lambda={v^\lambda}_{,\mu}+v^\rho\left\{{\lambda\atop \rho\mu}
\right\},\quad
\nabla_\mu v_\lambda=v_{\lambda,\mu}-v_\rho\left\{{\rho\atop \lambda\mu}
\right\},
\end{equation}
where $v^\mu$ is an arbitrary real vector.

(\ref{eq:2.22}) is equivalent to the set of equations:
\begin{equation}
\label{eq:3.3}
D_\sigma g_{\mu\nu}=2\Gamma^\rho_{[\sigma\nu]}g_{\mu\rho}
+L_{\mu\nu\sigma},
\end{equation}
where
\begin{equation}
L_{\mu\nu\sigma}=Y^\rho\Lambda_{\mu\nu\sigma\rho}.
\end{equation}
The system of equations (\ref{eq:3.3})
admits a solution of the form
\cite{bib:Hlavaty}:
\begin{equation}
\label{eq:3.5}
\Gamma^\lambda_{\mu\nu}=\left\{{\lambda\atop \mu\nu}\right\}
+K^\lambda_{(\mu\nu)}+\Gamma^\lambda_{[\mu\nu]},
\end{equation}
where
\begin{equation}
K^\lambda_{(\mu\nu)}=s^{\lambda\alpha}\Gamma^\beta_{[\alpha(\nu]}a_{\beta\mu)}
+\Omega^\lambda_{(\mu\nu)},
\end{equation}
\begin{equation}
\Omega^\lambda_{(\mu\nu)}
=\frac{1}{2}s^{\lambda\alpha}(L_{(\alpha\nu)\mu}+L_{(\alpha\mu)\nu}
-L_{(\mu\nu)\alpha}),
\end{equation}
and we have used the notation $a_{\mu\nu}=g_{[\mu\nu]}$.

The following identities contain the curvature tensor:
\begin{equation}
\label{eq:3.8}
2D_{[\nu} D_{\rho]}v^\lambda=-{R^\lambda}_{\mu\nu\rho}(\Gamma)v^\mu
+2\Gamma^\alpha_{[\nu\rho]}D_\alpha v^\lambda,
\end{equation}
\begin{equation}
2D_{[\nu}D_{\rho]}v_\mu={R^\lambda}_{\mu\nu\rho}(\Gamma)v_\lambda
+2\Gamma^\alpha_{[\nu\rho]}D_\alpha v_\mu,
\end{equation}
where
\begin{equation}
\label{eq:3.10}
{R^\lambda}_{\mu\nu\rho}(\Gamma)=\Gamma^\lambda_{\mu\nu,\rho}
-\Gamma^\lambda_{\mu\rho,\nu}-\Gamma^\lambda_{\alpha\nu}
\Gamma^\alpha_{\mu\rho}+\Gamma^\lambda_{\alpha\rho}
\Gamma^\alpha_{\mu\nu}.
\end{equation}
The curvature tensor satisfies the identities:
\begin{equation}
\label{eq:3.11}
{R^\lambda}_{\rho\mu\nu}(\Gamma)={R^\lambda}_{\rho[\mu\nu]}(\Gamma),
\end{equation}
\begin{equation}
\label{eq:3.12}
{R^\lambda}_{[\mu\nu\rho]}(\Gamma)
=4\Gamma^\alpha_{[[\nu\rho]}\Gamma^\lambda_{[\mu]\alpha]}
+2D_{[\nu}\Gamma^\lambda_{[\rho\mu]]},
\end{equation}
\begin{equation}
\label{eq:3.13}
D_{[\sigma} {R^\lambda}_{\vert\mu\vert\nu\rho]}(\Gamma)
=-2\Gamma^\alpha_{[[\sigma\nu]}{R^\lambda}_{\vert\mu\vert\rho]\alpha}(\Gamma).
\end{equation}
Let us define the two contracted curvature tensors:
\begin{equation}
W_{\mu\nu}={R^\beta}_{\mu\nu\beta}(\Gamma),\quad
V_{\nu\rho}={R^\beta}_{\beta\nu\rho}(\Gamma),
\end{equation}
where
\begin{equation}
W_{\mu\nu}=\Gamma^\beta_{\mu\nu,\beta}-\Gamma^\beta_{\mu\beta,\nu}
-\Gamma^\beta_{\alpha\nu}\Gamma^\alpha_{\mu\beta}+\Gamma^\beta_{\alpha\beta}
\Gamma^\alpha_{\mu\nu},
\end{equation}
and
\begin{equation}
V_{\nu\rho}=\Gamma^\beta_{\beta\nu,\rho}-\Gamma^\beta_{\beta\rho,\nu}.
\end{equation}
Then, from (\ref{eq:2.11}) and (\ref{eq:2.13}), we have
\begin{equation}
\label{eq:3.17}
R_{\mu\nu}(\Gamma)=W_{\mu\nu}+\frac{1}{2}(\Gamma^\beta_{(\mu\beta),\nu}
-\Gamma^\beta_{(\nu\beta),\mu}),
\end{equation}
and
\begin{equation}
V_{\nu\rho}=\Gamma^\beta_{(\beta\nu),\rho}-\Gamma^\beta_{(\beta\rho),\nu}.
\end{equation}

The Riemann-Christoffel curvature tensor ${B^\lambda}_{\mu\nu\rho}$
is defined by
\begin{equation}
{B^\lambda}_{\mu\nu\rho}=\left\{{\lambda\atop
\mu\nu}\right\}_{,\rho}-\left\{{\lambda\atop \mu\rho}\right\}_{,\nu}
-\left\{{\lambda\atop \alpha\nu}\right\}\left\{{\alpha\atop \mu\rho}\right\}
+\left\{{\lambda\atop \alpha\rho}\right\}\left\{{\alpha\atop \mu\nu}\right\}.
\end{equation}
By performing the contraction, ${B^\beta}_{\mu\nu\beta}=B_{\mu\nu}$, we get
\begin{equation}
\label{eq:3.20}
B_{\mu\nu}=\left\{{\alpha\atop \mu\nu}\right\}_{,\alpha}
-\left\{{\alpha\atop \mu\alpha}\right\}_{,\nu}
-\left\{{\beta\atop \alpha\nu}\right\}
\left\{{\alpha\atop \mu\beta}\right\}+\left\{{\beta\atop
\alpha\beta}\right\}\left\{{\alpha\atop \mu\nu}\right\}.
\end{equation}
The identities
(\ref{eq:3.11},\ref{eq:3.12},\ref{eq:3.13})
reduce for the tensor ${B^\lambda}_{\mu\nu\rho}$ to:
\begin{equation}
{B^\lambda}_{\rho\mu\nu}={B^\lambda}_{\rho[\mu\nu]},
\end{equation}
\begin{equation}
{B^\lambda}_{[\rho\mu\nu]}=0,
\end{equation}
\begin{equation}
\nabla_{[\sigma}{B^\lambda}_{\vert\mu\vert\nu\rho]}=0.
\end{equation}
The latter is the well-known Bianchi identity of Riemannian geometry.

Let us put
\begin{equation}
X_{\mu\nu}^\lambda=K_{(\mu\nu)}^\lambda+\Gamma_{[\mu\nu]}^\lambda.
\end{equation}
Then, we have from (\ref{eq:3.5}) and (\ref{eq:3.10}):
\begin{equation}
\Gamma^\lambda_{\mu\nu}=\left\{{\lambda\atop
\mu\nu}\right\}+X^\lambda_{\mu\nu},
\end{equation}
and
\begin{eqnarray}
{R^\lambda}_{\mu\nu\rho}(\Gamma) &=& {B^\lambda}_{\mu\nu\rho}
+X_{\mu\nu,\rho}^\lambda - X_{\mu\rho,\nu}^\lambda
-X_{\alpha\mu}^\lambda X^\alpha_{\nu\rho}
+X^\lambda_{\alpha\rho}X^\alpha_{\mu\nu} \nonumber \\
\label{eq:3.26} \mbox{}
& &\mbox{}-\left\{{\lambda\atop\alpha\nu}\right\}X^\alpha_{\mu\rho}-
\left\{{\alpha\atop\mu\rho}\right\}X^\lambda_{\alpha\nu}
+\left\{{\lambda\atop\alpha\rho}\right\}X^\alpha_{\mu\nu}
+\left\{{\alpha\atop\mu\nu}\right\}X^\lambda_{\alpha\rho}.
\end{eqnarray}
By contracting on $\lambda$ and $\rho$, we obtain from
(\ref{eq:2.13}) and (\ref{eq:3.26}):
\begin{eqnarray}
R_{\mu\nu}(\Gamma) &=& B_{\mu\nu}+X^\alpha_{\mu\nu,\alpha}
-{1\over 2}(X_{(\mu\alpha),\nu}^\alpha+X_{(\nu\alpha),\mu}^\alpha)
-X_{\alpha\nu}^\beta X_{\mu\beta}^\alpha
+X^\beta_{(\alpha\beta)}X^\alpha_{\mu\nu} \nonumber \\
& & \mbox{}
-\left\{{\beta\atop\alpha\nu}\right\}X^\alpha_{\mu\beta}-
\left\{{\alpha\atop\mu\beta}\right\}X^\beta_{\alpha\nu}
+\left\{{\beta\atop\alpha\beta}\right\}X^\alpha_{\mu\nu}
+\left\{{\alpha\atop\mu\nu}\right\}X^\beta_{(\alpha\beta)}.
\end{eqnarray}

\section{Geodesics, Paths and Path Deviation}

The line element is defined by
\begin{equation}
\label{eq:4.1}
d\tau^2=g_{\mu\nu}dx^\mu dx^\nu.
\end{equation}
Let us adopt the normalization condition:
\begin{equation}
\label{eq:4.2}
g_{\mu\nu}u^\mu u^\nu=1,
\end{equation}
where $u^\mu=dx^\mu/d\tau$ and $\tau$ is the proper time along the world
line of a particle.

A path is a curve in the spacetime, $(M, g_{\mu\nu})$, whose tangent $u^\mu$
is parallel transported along itself:
\begin{equation}
u^\sigma D_\sigma u^\mu=
{du^\mu\over d\tau}+\Gamma^\mu_{\alpha\beta}u^\alpha u^\beta=\phi(\tau)u^\mu.
\end{equation}
{}From (\ref{eq:4.2}), we have
\begin{equation}
\label{eq:4.4}
\frac{1}{2}D_\sigma(s_{\mu\nu}u^\mu u^\nu)
=\frac{1}{2}D_\sigma s_{\mu\nu}u^\mu u^\nu
+s_{\mu\nu}u^\nu D_\sigma u^\mu=0.
\end{equation}
Multipling (\ref{eq:4.4}) by $u^\sigma$ leads to the result:
\begin{equation}
u^\nu\left({du^\mu\over d\tau}
+\Gamma^\mu_{\alpha\beta}u^\alpha u^\beta\right)
s_{\mu\nu}=u^\nu\Sigma_\nu,
\end{equation}
where
\begin{equation}
\Sigma_\nu=-{1\over 2}u^\sigma D_\sigma s_{\mu\nu}u^\mu.
\end{equation}
{}From (\ref{eq:3.3}) it follows that
\begin{equation}
\Sigma_\nu=-\biggl(\Gamma^\rho_{[\sigma(\nu]}g_{\mu)\rho}
+\frac{1}{2}L_{(\mu\nu)\sigma}\biggr)u^\sigma u^\mu.
\end{equation}

When $u^\nu\Sigma_\nu=0$, we get $\phi(\tau)=0$ and the
path equation of motion
for test particles:
\begin{equation}
\label{eq:4.8}
\frac{du^\mu}{d\tau}+\Gamma^\mu_{\alpha\beta}u^\alpha u^\beta=0.
\end{equation}

{}From the action:
\begin{equation}
I=\int d\tau^2,
\end{equation}
where $d\tau^2$ is given by (\ref{eq:4.1}),
we can derive from Fermat's principle
the geodesic equation of motion:
\begin{equation}
\frac{du^\mu}{d\tau}
+\left\{{\mu\atop \alpha\beta}\right\}u^\alpha u^\beta=0,
\end{equation}
which is equivalent to the equation:
\begin{equation}
u^\sigma\nabla_\sigma u^\mu=0.
\end{equation}

The equation $u^\nu\Sigma_\nu=0$ will not in general be satisfied in
NGT. However, we are interested in studying the path completeness of
the NGT spacetime, so in the following we shall adopt
(\ref{eq:4.8}) as an equation
of motion of particles that probe the regularity properties of
the spacetime geometry.

The geodesic equation of motion is not equivalent to the path equation,
because in general:
\begin{equation}
\left(\Gamma^\mu_{\alpha\beta}-\left\{{\mu\atop \alpha\beta}\right\}\right)
u^\alpha u^\beta \not=0,
\end{equation}
although $g_{\mu\nu}u^\mu u^\nu$ is a constant of the motion for both curves.

Let us define the torsion by
\begin{equation}
T(A,B)=D_A B-D_B A-[A,B],
\end{equation}
which is equivalent to
\begin{equation}
T(A,B)=2A^\mu B^\nu\Gamma^\alpha_{[\mu\nu]}e_\alpha,
\end{equation}
where $A^\mu$ and $B^\mu$ are two arbitrary real vectors and $e_\alpha$
is a basis vector.

Consider a congruence of curves with tangent vectors $V^\mu$. The separation
vector $\eta^\mu$ of two nearby curves is a vector which connects them at
equal values of their curve parameter $\lambda$. The rate of change of
$\eta^\mu$ along the congruence is zero:
\begin{equation}
[V,\eta]=(V^\alpha\partial_\alpha \eta^\beta
- \eta^\alpha\partial_\alpha V^\beta)e_\beta=0.
\end{equation}

The geodesic deviation equation is defined by
\begin{equation}
\nabla_V\nabla_V\,\eta=B(V,\eta)V+\nabla_\eta(\nabla_V\, V),
\end{equation}
where
\begin{equation}
B(V,\eta)V=\nabla_V\nabla_\eta V-\nabla_\eta\nabla_V V - \nabla_{[V,\eta]}V.
\end{equation}
The path deviation equation is defined by
\begin{equation}
\label{eq:4.18}
D_V D_V\,\eta=R(V,\eta)V+D_V[T(V,\eta)]+D_\eta(D_V V),
\end{equation}
where
\begin{equation}
R(V,\eta)V=D_V D_\eta V-D_\eta D_V V - D_{[V,\eta]}V.
\end{equation}
In general the geodesic and path deviation equations are not equivalent
in the spacetime $(M, g_{\mu\nu})$.

\section{Static Spherically Symmetric Solution}

In the case of a static spherically symmetric field, the canonical form of
$g_{\mu\nu}$ is given by:
\begin{equation}
g_{\mu\nu}=\left(
\begin{array}{cccc}
-\alpha&0&0&w \\
0&-\beta&f\sin\theta&0 \\ 0&-f\sin\theta&-\beta\sin^2
\theta&0\cr-w&0&0&\gamma\end{array}\right),
\end{equation}
where $\alpha,\beta,\gamma$ and $w$ are functions of $r$. The
tensor $g^{\mu\nu}$ has the components:
\begin{equation}
g^{\mu\nu}=\left(\begin{array}{cccc}
\frac{\gamma}{w^2-\alpha\gamma}&0&0&\frac{w}{w^2-\alpha\gamma} \\
0&-\frac{\beta}{\beta^2+f^2}&\frac{f\csc\theta}{\beta^2+f^2}&0 \\
0&-\frac{f\csc\theta}{\beta^2+f^2}
&-\frac{\beta\csc^2\theta}{\beta^2+f^2}&0\\
-\frac{w}{w^2-\alpha\gamma}&0&0&-\frac{\alpha}{w^2-\alpha\gamma}\\
\end{array}\right).
\end{equation}
For the theory in which there is no NGT magnetic monopole charge, we have
$w=0$ and only the $g_{[23]}$ component of $g_{[\mu\nu]}$ survives.
The line element for a spherically symmetric body is given by
\begin{equation}
ds^2=\gamma(r)dt^2-\alpha(r)dr^2-\beta(r)(d\theta^2+\sin^2\theta
d\phi^2).
\end{equation}
We have
\begin{equation}
\sqrt{-g}=\sin\theta[\alpha\gamma(\beta^2+f^2)]^{1/2}.
\end{equation}
For the static spherically symmetric field with $w=0$, it follows from
(\ref{eq:2.20})
that $W_\mu=0$.

Let us assume the long-range approximation for which the $\mu^2$ contributions
in the vacuum field equations can be neglected and we assume that $\Lambda=0$.
We then obtain the static, spherically symmetric Wyman solution
\cite{bib:Cornish_1,bib:Cornish_2}:
\begin{equation}
\gamma=\exp(\nu),
\end{equation}
\begin{equation}
\alpha=m^2(\nu')^2\exp(-\nu)(1+s^2)
(\cosh(a\nu)-\cos(b\nu))^{-2},
\end{equation}
\begin{equation}
f=[2m^2\exp(-\nu)(\sinh(a\nu)\sin(b\nu)+s(1-\cosh(a\nu)
\cos(b\nu))](\cosh(a\nu)-\cos(b\nu))^{-2},
\end{equation}
where
\begin{equation}
a=\left(\frac{\sqrt{1+s^2}+1}{2}\right)^{1/2},\quad
b=\left(\frac{\sqrt{1+s^2}-1}{2}\right)^{1/2},
\end{equation}
prime denotes differentiation with respect to $r$, and $\nu$ is implicitly
determined by the equation:
\begin{equation}
\exp(\nu)(\cosh(a\nu)-\cos(b\nu))^2\frac{r^2}{2m^2}=
\cosh(a\nu)\cos(b\nu)-1+s\sinh(a\nu)\sin(b\nu).
\end{equation}
Moreover, $s$ is a dimensionless constant of integration.

We find for $2m/r \ll 1$ and $0 < sm^2/r^2 < 1$ that the metric
takes the Schwarzschild form:
\begin{equation}
\gamma\sim\alpha^{-1}\sim1-\frac{2m}{r},
\end{equation}
and
\begin{equation}
f\sim\frac{sm^2}{3}.
\end{equation}
Near $r=0$ we can develop expansions where $r/m < 1$ and $0 < \vert s\vert
<1$. The leading terms are
\cite{bib:Cornish_1,bib:Cornish_2}:
\begin{equation}
\gamma=\gamma_0+
\frac{\gamma_0(1+{\cal O}(s^2))}{2\vert s\vert}\left(\frac{r}{m}\right)^2
+ {\cal O}\left(\left(\frac{r}{m}\right)^4\right),
\end{equation}
\begin{equation}
\alpha=\frac{4\gamma_0(1+{\cal O}(s^2))}{s^2}\left(\frac{r}{m}\right)^2
+{\cal O}\left(\left(\frac{r}{m}\right)^4\right),
\end{equation}
\begin{equation}
f=m^2\left(4-\frac{\vert s\vert\pi}{2}+s\vert s\vert+{\cal O}(s^3)\right)
+\frac{\vert s\vert+s^2\pi/8+{\cal O}(s^3)}{4}r^2+{\cal O}(r^4),
\end{equation}
\begin{equation}
\gamma_0=\exp\left(-\frac{\pi+2s}{\vert s\vert}+{\cal O}(s)\right)
\ldots \, .
\end{equation}

These solutions clearly illustrate
the non-analytic nature of the limit $s\rightarrow 0$ in the strong
gravitational field regime for $r < 2m$.

The singularity caused by the vanishing of $\alpha(r)$ at
$r=0$ is a {\it coordinate} singularity, which can be removed by
transforming to another coordinate frame of reference. The curvature
invariants do not, of course, contain any coordinate singularities.
We can transform to a coordinate frame in which the spacetime
$(M,g_{\mu\nu})$ is completely free of singularities
\cite{bib:Cornish_1,bib:Cornish_2}
The NGT curvature invariants such as the generalized Kretschmann scalar:
\begin{equation}
K=R^{\lambda\mu\nu\rho}R_{\lambda\mu\nu\rho}
\end{equation}
are finite. However, the curvature invariants formed from the
Riemann-Christoffel tensor are singular at $r=0$. For example, the Kretschmann
scalar:
\begin{equation}
S=B^{\lambda\mu\nu\rho}B_{\lambda\mu\nu\rho},
\end{equation}
is singular like $S\sim m^4/r^8$ near $r=0$
\cite{bib:Cornish_3}.
The
Christoffel symbol defined by (\ref{eq:2.29})
is singular at $r=0$, and shares the
same analytic spacetime properties as the Christoffel symbol in EGT.
Thus, only
the fully non-Riemannian geometry of NGT describes a nonsingular spacetime.

The static spherically symmetric solution is everywhere non-singular and
there is no event horizon at $r=2m$. A black hole is replaced in the theory
by a FCO which can be stable for an arbitrarily large mass
\cite{bib:Cornish_1,bib:Cornish_2,bib:Cornish_3}.
It has been shown by Cornish \cite{bib:Cornish_4}, that the static
spherically symmetric solution of the field equations for $\mu\not=0$
only satisfies the flat space asymptotic conditions at $r\rightarrow\infty$
if $w=0$. This demonstrates that the Wyman solution is the
unique static spherically symmetric solution of the field equations and
that only one degree of freedom is permitted globally for this solution.

Because the Christoffel symbol $\left\{{\mu\atop
\alpha\beta}\right\}$ is singular at $r=0$ in the spacetime, $(M, g_{\mu\nu})$,
it follows that the Wyman solution is not geodesically complete. On the
other hand, the connection $\Gamma^\lambda_{\mu\nu}$ is non-singular
everywhere in the spacetime, $(M, g_{\mu\nu})$, including at $r=0$, so
the path equation (\ref{eq:4.8})
can be complete for the Wyman solution. Because
$Y^\mu=0$ and therefore $u^\nu\Sigma_\nu=0$ for the Wyman solution, it follows
that the path equation (\ref{eq:4.8}) holds for this solution.

\section{Path Congruences}

Let a smooth congruence of timelike paths be parameterized by proper
time $\tau$, so that the vector field, $V^\mu$, of tangents is normalized to
unit length:
\begin{equation}
V^\mu V_\mu=s_{\lambda\mu}V^\lambda V^\mu=1.
\end{equation}
We define a tensor field $Q_{\mu\nu}$ by
\begin{equation}
Q_{\mu\nu}=D_\nu V_\mu.
\end{equation}
It is purely spatial in character:
\begin{equation}
Q_{\mu\nu}V^\mu=Q_{\mu\nu}V^\nu=0.
\end{equation}
Let $\gamma$ denote a smooth one-parameter subfamily of paths in the
congruence, and let $\eta^\mu$ describe an infinitesimal displacement
from one such path $\gamma_0$ to a nearby path in the subfamily.
Then, we have
\begin{equation}
V^\nu D_\nu \eta^\mu=\eta^\nu D_\nu V^\mu={Q^\mu}_\nu \eta^\nu.
\end{equation}

Let us define a ``spatial'' metric tensor $h_{\mu\nu}$:
\begin{equation}
h_{\mu\nu}=s_{\mu\nu}-V_\mu V_\nu,
\end{equation}
which satisfies $h_{\mu\nu}V^\mu=h_{\mu\nu}V^\nu=0$ and ${h_\mu}^\nu
=s^{\nu\sigma}h_{\mu\sigma}$. Then, the expansion $\theta$, shear
$\sigma_{\mu\nu}$, and twist
$\omega_{\mu\nu}$ of the congruence are given, respectively, by
\begin{equation}
\theta=Q^{\mu\nu}h_{\mu\nu},
\end{equation}
\begin{equation}
\sigma_{\mu\nu}=Q_{(\mu\nu)}-{1\over 3}\theta h_{\mu\nu},
\end{equation}
\begin{equation}
\omega_{\mu\nu}=Q_{[\mu\nu]}.
\end{equation}
We have
\begin{equation}
Q_{\mu\nu}=\frac{1}{3}\theta h_{\mu\nu}+\sigma_{\mu\nu}+\omega_{\mu\nu},
\end{equation}
and $\sigma_{\mu\nu}V^\mu=\omega_{\mu\nu}V^\nu=0$.

It can be shown from (\ref{eq:3.8}) that
\begin{equation}
V^\sigma D_\sigma {Q^\lambda}_\nu=-{Q^\sigma}_\nu {Q^\lambda}_\sigma
-{R^\lambda}_{\mu\sigma\nu}(\Gamma) V^\mu V^\sigma
+ 2\Gamma^\alpha_{[\sigma\nu]}{Q^\lambda}_\alpha V^\sigma,
\end{equation}
Contracting on the suffixes $\lambda$ and $\nu$, we get the generalized
Raychaudhuri equation \cite{bib:Raychaudhuri}:
\begin{equation}
\label{eq:6.11}
\frac{d\theta}{d\tau}=-\frac{1}{3}\theta^2-\sigma^2+\omega^2
- R_{\mu\nu}(\Gamma)V^\mu V^\nu
- 2\Gamma^\alpha_{[\mu\nu]}{Q^\mu}_\alpha V^\nu,
\end{equation}
where we have used $W_{\mu\nu}V^\mu V^\nu=R_{\mu\nu}(\Gamma)V^\mu V^\nu$,
which follows from (\ref{eq:3.17}). Moreover,
\begin{equation}
\sigma^2=h_{\mu\alpha}h_{\nu\beta}\sigma^{\alpha\beta}\sigma^{\mu\nu},\quad
\omega^2=h_{\mu\alpha}h_{\nu\beta}\omega^{\alpha\beta}\omega^{\mu\nu}.
\end{equation}

Next we consider a congruence of null paths with tangent field $Z^\mu$.
We associate the metric ${\hat h}_{\mu\nu}={\hat h}_{\nu\mu}$ with the space
of null vectors $Z^\mu$ and we define the tensor field:
\begin{equation}
{\hat Q}_{\mu\nu}=D_\nu Z_\mu,
\end{equation}
which can be decomposed as:
\begin{equation}
{\hat Q}_{\mu\nu}=\frac{1}{2}{\hat \theta}{\hat h}_{\mu\nu}
+{\hat \sigma}_{\mu\nu}+{\hat \omega}_{\mu\nu},
\end{equation}
where, as before, ${\hat \theta}, {\hat \sigma}_{\mu\nu}$ and
${\hat \omega}_{\mu\nu}$ denote the expansion, shear and twist, respectively,
given by
\begin{equation}
{\hat \theta}={\hat Q}^{\mu\nu}{\hat h}_{\mu\nu},
\end{equation}
\begin{equation}
{\hat \sigma}_{\mu\nu}={\hat Q}_{(\mu\nu)}-\frac{1}{2}{\hat \theta}
{\hat h}_{\mu\nu},
\end{equation}
\begin{equation}
{\hat \omega}_{\mu\nu}={\hat Q}_{[\mu\nu]}.
\end{equation}
Then, for an affine parameter $\lambda$, the generalized Raychaudhuri
equation takes the form:
\begin{equation}
\label{eq:6.18}
\frac{d{\hat \theta}}{d\lambda}=-\frac{1}{2}{\hat \theta}^2
-{\hat \sigma}^2+{\hat \omega}^2-R_{\mu\nu}(\Gamma)Z^\mu Z^\nu
- 2\Gamma^\alpha_{[\mu\nu]}{{\hat Q}^\mu}_\alpha Z^\nu.
\end{equation}

In Einstein's theory of gravity, we have
\begin{equation}
a_{\mu\nu}=0,\quad \Gamma^\lambda_{[\mu\nu]}=0,
\end{equation}
and (\ref{eq:3.3}) becomes
\begin{equation}
\nabla_\sigma s_{\mu\nu}=0.
\end{equation}
Einstein's gravitational field equations give:
\begin{equation}
B_{\mu\nu}V^\mu V^\nu=8\pi(T_{(\mu\nu)}-\frac{1}{2}s_{\mu\nu}T)V^\mu V^\nu
=8\pi(T_{(\mu\nu)} V^\mu V^\nu-\frac{1}{2}T).
\end{equation}
For physical matter we have
\begin{equation}
\label{eq:6.22}
T_{(\mu\nu)}V^\mu V^\nu \geq 0,
\end{equation}
for all timelike vectors $V^\mu$, which is called the weak energy condition.
Moreover,
\begin{equation}
\label{eq:6.23}
T_{(\mu\nu)}V^\mu V^\nu \geq \frac{1}{2}T,
\end{equation}
for timelike $V^\mu$ is the strong energy condition.

In the case of null geodesics, we obtain from Einstein's field equations:
\begin{equation}
\label{eq:6.24}
B_{\mu\nu}=8\pi T_{(\mu\nu)}Z^\mu Z^\nu,
\end{equation}
and we postulate that
\begin{equation}
\label{eq:6.25}
T_{(\mu\nu)}Z^\mu Z^\nu \geq 0,
\end{equation}
which is a weaker requirement than for the timelike vectors $V^\mu$.

For Einstein's theory, (\ref{eq:6.11}) and (\ref{eq:6.18}) become
\begin{equation}
\label{eq:6.26}
\frac{d\theta}{d\tau}=-\frac{1}{3}\theta^2-\sigma^2+\omega^2
-B_{\mu\nu}V^\mu V^\nu,
\end{equation}
\begin{equation}
\label{eq:6.27}
\frac{d{\hat \theta}}{d\lambda}=-\frac{1}{2}{\hat \theta}^2
-{\hat \sigma}^2+{\hat \omega}^2-B_{\mu\nu}Z^\mu Z^\nu.
\end{equation}
{}From (\ref{eq:6.22}) and (\ref{eq:6.25}),
it follows that the last terms of (6.26) and (6.27)
are negative. The
terms $\sigma^2$ and ${\hat \sigma}^2$,
in (\ref{eq:6.26}) and (\ref{eq:6.27}), are nonpositive
and if the congruence of geodesics is hypersurface orthogonal, we have
$\omega_{\mu\nu}={\hat \omega}_{\mu\nu}=0$, so that the third terms in
(\ref{eq:6.26}) and (\ref{eq:6.27}) vanish.
If we set $z=\theta^{-1}$ and ${\dot z}=dz/d\tau$, then we get from
(\ref{eq:6.26}):
\begin{equation}
{\dot z} \geq \frac{1}{3},
\end{equation}
or,
\begin{equation}
\label{eq:6.29}
z(\tau) \geq z_0^{-1}+\frac{1}{3}\tau,
\end{equation}
where $z_0$ is the initial value of $z$. If $\theta_0$ is negative,
corresponding to an initially converging geodesic congruence, then
(\ref{eq:6.29}) requires
that $z$ must pass through zero, which means that $\theta\rightarrow
-\infty$ within a proper time $\tau \leq 3/\vert\theta_0\vert$. Unboundedness
of $\theta$ implies that the volume of a tube of matter shrinks to zero
and for normal matter, $\rho\rightarrow\infty$ and $p\rightarrow\infty$,
generating a singularity.

In NGT, the symmetric part of the field equations (\ref{eq:2.25}) is given by
\begin{equation}
S_{(\mu\nu)}V^\mu V^\nu=8\pi(T_{(\mu\nu)}V^\mu V^\nu-\frac{1}{2}T).
\end{equation}
As in EGT, we assume for physical matter that (\ref{eq:6.22})
and (\ref{eq:6.23}) hold.

Let us write
\begin{equation}
X(\tau)=R_{\mu\nu}(\Gamma)V^\mu V^\nu
+\sigma^2 +2\Gamma^\alpha_{[\mu\nu]}{Q^\mu}_\alpha V^\nu.
\end{equation}
For a hypersurface orthogonal congruence of paths,
$\omega_{\mu\nu}=0$, and (\ref{eq:6.11}) can be written as
\begin{equation}
{\dot z}-X(\tau)z^2-\frac{1}{3}=0,
\end{equation}
which has the form of a Riccati equation. We have
\begin{equation}
X(\tau)=\frac{{\dot z}-\frac{1}{3}}{z^2}.
\end{equation}
If $X(\tau)$ is a smooth analytic function of $\tau$, then $z\not=0$ for all
$\tau$. We exclude the exceptional case when
${\dot z}\rightarrow \frac{1}{3}$ as $z^2$ for some $\tau=\tau_0$.

We have proven the following Lemma:
\begin{lemma}
\label{lemma:6.1}
Let $V^\mu$ be the tangent field of a hypersurface orthogonal
timelike path congruence. Suppose that $X(\tau)$ is a smooth analytic
function of $\tau$. Then, $\vert\theta\vert < \infty$ along a path in that
congruence for any $\tau$ and the path can be extended indefinitely.
\end{lemma}
For a congruence of null paths with tangent field $Z^\mu$, we write
\begin{equation}
{\hat X}(\tau)=R_{\mu\nu}(\Gamma)Z^\mu Z^\nu+{\hat \sigma}^2
+ 2\Gamma^\alpha_{[\mu\nu]}{{\hat Q}^\mu}_\alpha Z^\nu,
\end{equation}
and we obtain from
(\ref{eq:6.18}) for a hypersurface orthogonal congruence of paths:
\begin{equation}
{\hat X}(\tau)=\frac{dz/d\lambda - \frac{1}{2}}{z^2}.
\end{equation}
We can state the following Lemma:
\begin{lemma}
\label{lemma:6.2}
Let $Z^\mu$ be the tangent field of a hypersurface orthogonal null
path congruence. Let us suppose that ${\hat X}(\tau)$ is a smooth analytic
function of $\tau$. Then,
for a path in the congruence, $\vert\theta\vert < \infty$
along that path for any affine length $\lambda$, and the null path
can be extended indefinitely.
\end{lemma}
A Jacobi field on a path $\gamma$ with tangent $V^\mu$ is a solution of
the path deviation equation (\ref{eq:4.18}).
Points $a,b\in \gamma$ are conjugate if there exists a field $\eta^\mu$ which
vanishes at both $a$ and $b$. Conjugate points in spacetime signal the
moment when a curve fails to be a local maximum of proper
time between two points and a null path fails to remain on
the boundary of the future of a point. In Riemannian geometry, they
signal locally the moment when a curve fails to take on its minimal length.
We can now state the following
\cite{bib:Hawking_3,bib:Wald}:
\begin{proposition}
\label{prop:6.1}
Let $(M,g_{\mu\nu})$ be a strongly causal spacetime, and
let $a,b\in M$ with
$b\in J^+(b)$, then the length function $\tau$, defined on a Cauchy
surface $\Sigma(p)$, attains a finite value at $\gamma\in \Sigma(p)$
when $\gamma$ is a path orthogonal to $\Sigma$ with no point conjugate
to $\sigma$ between $\Sigma$ and $a$.
\end{proposition}
The extrinsic curvature, $K_{\mu\nu}$, on a spacelike hypersurface
$\Sigma$ is defined by
\begin{equation}
K_{\mu\nu}=\nabla_\mu V_\nu=Q_{\mu\nu}.
\end{equation}
$K_{\mu\nu}$ is purely spatial i.e., $K_{\mu\nu}V^\mu=K_{\mu\nu}V^\nu=0$.
The trace of $K_{\mu\nu}$ is given by
\begin{equation}
K=K^{\mu\nu}h_{\mu\nu}=\theta,
\end{equation}
where $\theta$ is the expansion of the path congruence orthogonal to $\Sigma$.

\section{Regularity Theorems for Cosmology}\par

When a spacetime manifests timelike or null geodesic incompleteness,
then this has the immediate physical significance that there are
freely falling observers whose histories did not exist after, or before,
a finite interval of proper time. Thus, timelike and null geodesic
completeness are minimum conditions for a singularity-free spacetime.
A spacetime which is null or timelike geodesically incomplete
contains a singularity. The notion of geodesic completeness can be
extended to bundle-completeness (b-completeness) i.e., that a spacetime
$(M,g_{\mu\nu})$ is b-complete if there is an endpoint for every
$C^1$ curve of finite length as measured by a generalized affine parameter.
Thus, a spacetime is singularity-free if it is b-complete
\cite{bib:Schmidt}.
For non-Riemannian gravitational theories these statements can be
extended to autoparallel paths of the connection $\Gamma^\lambda_{\mu\nu}$.

Let us consider proving two theorems which establish the necessary
conditions for the existence of nonsingular solutions in NGT cosmology.
We shall use the properties of the non-Riemannian geometry of NGT,
whose associated timelike and null paths reveal under what conditions
these paths can be complete.

We shall first assume that the universe is globally hyperbolic and that at
one instant of time it is expanding everywhere at a nonzero rate. Then
the universe will have begun in a nonsingular state a finite time ago
if:
\begin{theorem}
Let $(M,g_{\mu\nu})$ be a globally hyperbolic spacetime in which
the strong energy condition, $S_{(\mu\nu)}V^\mu V^\nu \geq 0$, holds and
$X(\tau)$
is a smooth analytic function of $\tau$ for all timelike $V^\mu$.
Assume that there exists a smooth spacelike Cauchy surface $\Sigma$ for which
the trace of the extrinsic curvature for the past directed normal path
congruence satisfies $K \leq C < 0$ everywhere (where $C$ is a constant).
Then all timelike paths from $\Sigma$ are complete.
\end{theorem}
\noindent
{\it Proof}. Let there exist a past timelike directed curve, $\gamma$,
from $\Sigma$. Let $a$ be a point
on $\gamma$ lying beyond length $3/\vert C\vert$ from $\Sigma$.
By Lemma \ref{lemma:6.1}
and Proposition \ref{prop:6.1}, there exists a curve $\gamma$ from $a$ to
$\Sigma$, which also must have length greater than $3/\vert C\vert$.
Since $\vert\theta\vert < \infty$ at all points on $\gamma$, then $\gamma$
is a path with no conjugate point between $\Sigma$ and $a$ and
therefore the path is complete.
\hfill$\Box$

We can eliminate the assumption of global hyperbolicity in the same way
that is done in the proof of the Hawking-Penrose theorem
\cite{bib:Hawking_2}, which
establishes the initial singularity in EGT cosmology,
by assuming that $\Sigma$
is compact (the universe is closed).
\begin{theorem}
Suppose that $S_{(\mu\nu)}N^\mu N^\nu \geq 0$, which
follows from the strong energy condition for timelike and null vectors $N^\mu$.
Let $(M,g_{\mu\nu})$ be a strongly
causal spacetime such that $X(\tau)$ and ${\hat X}(\tau)$ are smooth analytic
functions of $\tau$. Let there exist
a compact, edgeless, achronal, smooth spacelike hypersurface $\Gamma$ such that
the past directed normal path congruence from $\Gamma$ has $K < 0$
everywhere on $\Gamma$. Then all the past directed timelike and null paths
can be extended beyond the value $3/\vert C\vert$ (where $C$ denotes
the maximum value of $K$, so that $K \leq C < 0$) and therefore the
paths are complete.
\end{theorem}
\noindent
{\it Proof}. According to Lemmas \ref{lemma:6.1}
and \ref{lemma:6.2} all timelike
and null paths past directed from $\Gamma$ will have
$\vert\theta\vert < \infty$ at all points on the paths, which means that
there will be no conjugate points on the paths beyond the value
$3/\vert C\vert$. Then these paths can be extended indefinitely.
\hfill$\Box$

Let us define
\begin{equation}
Y(\tau)=B_{\mu\nu}V^\mu V^\nu + \sigma^2,
\end{equation}
and
\begin{equation}
{\hat Y}(\tau)=B_{\mu\nu}Z^\mu Z^\nu + {\hat \sigma}^2,
\end{equation}
where $B_{\mu\nu}$ is given by (\ref{eq:3.20}).

In Einstein's gravitational theory, the Hawking-Penrose theorems establish that
if the positive energy condition,
$B_{\mu\nu}N^\mu N^\nu \geq 0$, holds for timelike and null vectors $N^\mu$,
then at least one past directed timelike
geodesic has no length greater than $3/\vert C\vert$, and is therefore
inextendible. This means that $z$ must vanish at some point on the geodesic
and, therefore, $Y(\tau)$ and ${\hat Y}(\tau)$ cannot be smooth analytic
functions for all values of the proper time or affine parameter. The physical
interpretation of the
positive energy condition is that gravity is always attractive i.e.,
neighboring geodesics near any one point accelerate, on the average, toward
each other.

In NGT, we can prove that all paths
are complete provided that for hypersurface normal
congruences, $X(\tau)$ and ${\hat X}(\tau)$ are smooth analytic functions
of $\tau$. This is now possible because the positive energy condition,
$S_{(\mu\nu)}N^\mu N^\nu\geq 0$, does not conflict with the analytic
property of $X(\tau)$ or ${\hat X}(\tau)$. The functions $X(\tau)$ and
${\hat X}(\tau)$
can be positive or negative in the spacetime manifold without violating
the positive energy condition.
Thus, NGT belongs to a class of non-Riemannian
theories in which gravity is not always attractive, although physical
matter satisfies the required positive energy conditions. This is why
singular curvature can be avoided in gravitationally unstable situations.
It is clear that any gravity theory based on a purely Riemannian geometry
(excluding $R^2$ curvature terms in the Lagrangian density) must
have incomplete geodesics and that all cosmological solutions begin
their expansion in a singular state. Any
external fields in such a theory of gravity will be incorporated in the
positive energy conditions
and inevitably lead to the Hawking-Penrose theorems. An example of this
situation is provided by the de Sitter solution in EGT cosmology for
the equation of state: $\rho=-p$. The solution is nonsingular, but violates
the positive energy condition.
On the other hand, NGT is based on a non-Riemannian geometry which
allows nonsingular solutions in cosmology to exist.

\section{Regularity Theorem for Stellar Collapse}

The solution of the NGT field equations outside a star is necessarily that
part of the asymptotically flat region of the Wyman solution for
which $r$ is greater than some value $R$ corresponding to the radius of
the star. For a given pressure and density in the star a solution
will exist, which is joined, for $r > R$, onto the exterior solution.
Since time-like
Killing vectors remain time-like for all $r$ between $r=0$ and $r=\infty$,
the star can have $r < 2m$. This is in contrast to the Schwarzschild
solution in EGT, for which the star must have $r > 2m$, because only in this
case is there a time-like Killing vector.
Let us now suppose that the nuclear fuel of the star is exhausted,
causing the star to contract under the influence of its gravity.
If the contraction cannot be halted by the pressure before $r \leq 2m$,
due to the star having a mass greater than some critical mass $m_c$, and if
the collapse ends in a final static spherically symmetric configuration, then
the solution outside the star is the nonsingular Wyman solution, and
we can anticipate that the final collapsed star has no event horizon and no
singularity at $r=0$.

In the non-singular spacetime, $(M, g_{\mu\nu})$,
the entire causal past of future null infinity,
$J^-({\cal J}^+)$, is nonsingular, and includes the entire physical spacetime.
Thus, there are no {\it black hole regions}, ${\cal R}_B$, defined by
${\cal R}_B=[M-J^-({\cal J^+})]$ with an enclosing event horizon
boundary: $H={\dot J}^-({\cal J}^+)\cap  M$.
Let there be some past-directed unit timelike vector $V^\mu$
at a point $p$. A compact, two-dimensional, smooth
spacelike submanifold, ${\cal T}$, for which the expansion, $\theta$,
of both sets of ingoing and outgoing future directed null geodesics or paths
orthogonal to ${\cal T}$ is everywhere negative is called a trapped
surface. In the Schwarzschild solution, all spheres inside the black hole
($r < 2m$) are trapped surfaces. In the nonsingular Wyman solution, there
will not be any trapped surfaces ${\cal T}$ below $r=2m$, since the $\theta$
associated with the paths is everywhere non-negative.
We can now state the following regularity theorem:
\begin{theorem}
\label{theorem:8.1}
The spacetime $(M, g_{\mu\nu})$ can be path complete if:
\begin{enumerate}
\item Let $(M, g_{\mu\nu})$ be a spacetime which satisfies the strong energy
condition for matter: $S_{(\mu\nu)}N^\mu N^\nu\geq 0$ for all timelike and null
vectors $N^\mu$.
\item $X(\tau)$ and ${\hat X}(\tau)$ are smooth analytic functions of
$\tau$ for
every non-spacelike vector $N^\mu$.
\item There are no closed time-like curves.
\item There are no closed trapped surfaces ${\cal T}$.
\end{enumerate}
\end{theorem}
\noindent
{\it Proof}. Let $\theta_0 < 0$ denote the maximum value of $\theta$ for
a set of path congruences in $(M, g_{\mu\nu})$. Because
there are no trapped surfaces in the spacetime all paths
have affine length $\geq 3/\vert\theta_0\vert$ ($\geq 2/\vert\theta_0\vert$
for null paths). The smooth analytic
behavior of $X(\tau)$ and ${\hat X}(\tau)$ guarantees by Lemmas
\ref{lemma:6.1} and \ref{lemma:6.2}
that $z\not=0$ along
the paths for $\tau\geq 3/\vert\theta_0\vert$ (or $2/\vert\theta_0\vert$)
and by Proposition \ref{prop:6.1}, the paths will not have any
conjugate points at
which the paths terminate. Therefore, the paths are null and timelike
complete.
\hfill$\Box$

In the nonsingular Wyman solution, $X(\tau)$ and ${\hat X}(\tau)$ are smooth
analytic functions of $\tau$ for all timelike and null vectors $N^\mu$. It
follows that this solution satisfies the
conditions of Theorem \ref{theorem:8.1}.
Gravity
is not always attractive, for $R_{\mu\nu}(\Gamma)N^\mu N^\nu$ can change sign
and can be negative as the gravitational field increases its strength, so that
neighboring paths will accelerate, on the average, away from each other.
Even if the star is not spherically symmetric, there will still not be any
trapped surface provided the departures from spherical symmetry are not too
large. This will follow from the stability of the Cauchy development,
since trapped surfaces cannot form in any gravitational collapse whose
initial conditions are sufficiently close to initial conditions for
spherical collapse. These results can be generalized
to a whole manifold, and the existence and uniqueness of developments for an
initial set of data can be derived. As in EGT, there exist constraint
equations whose development and uniqueness for an initial set of data
can be proved rigorously. The
analog of the stability proof for small fluctuations of the spherically
symmetric Schwarzschild solution
\cite{bib:Vishveshwara}
has not yet been established for the
nonsingular Wyman solution. However, since the theory has a stable vacuum
solution that follows from a consistent perturbative expansion scheme,
we expect that the NGT static spherically symmetric solution is stable
against small fluctuations.

{}From the static Wyman solution, we find
that $Y(\tau)$ is not an analytic smooth function everywhere in the spacetime
$(M, g_{\mu\nu})$, so the Riemannian geometry associated with the NGT spacetime
does not satisfy the regularity theorems established in the foregoing, and the
geodesics in $(M, g_{\mu\nu})$ are not complete.

\section{Conclusions}

When the physical
matter satisfies positive and non-negative pressure conditions and
$g_{\mu\nu}
=s_{\mu\nu}$, the spacetime possesses a Riemannian geometry and if the theory
is of Einstein form i.e., the Lagrangian
density contains no derivatives higher than the second order, then the
Hawking-Penrose theorems are valid and the spacetime must contain
non-coordinate singularities. Classical alternatives to this singular
spacetime scenario are a gravity theory based on non-Riemannian geometry
and a theory based on Riemannian geometry and a higher-derivative Lagrangian
density. In particular, we have studied NGT and found that it
can satisfy the regularity theorems for cosmology and gravitational collapse
proved above.

We have argued that any gravity theory with a set of initial data on a Cauchy
surface that has a global non-perturbative solution to the field
equations, which can be mapped into solutions near the Schwarzschild solution
of EGT, shares the singularity at $r=0$ and the black
hole event horizon of the Schwarzschild solution. For this reason, we must
expect that any modified gravity theory that removes all singularities and
black hole event horizons is non-perturbative and non-analytic in the
new coupling constants for $r < 2m$.

The static spherically symmetric Wyman solution of the vacuum field equations
is non-singular in the long-range approximation and
satisfies the conditions of the regularity theorems, namely, the spacetime is
path complete and the curvature invariants are finite. The solution is
not analytic in the parameter $s$
for $r \leq 2m$, for in the limit $s\rightarrow 0$ the solution does not
become the Schwarzschild solution with its event horizon and singularity at
$r=0$.
Since the solution does not have any event horizons there are no black holes.
The black hole is replaced by a final collapsed object that can be made stable
by the repulsive NGT forces for an arbitrarily large mass.

An observer, in EGT, falling through an event horizon of a large black hole
does not experience anything unusual about the event, whereas in NGT the
observer would in general detect a skew symmetric force at the event horizon,
enabling him to determine the existence of such a horizon.

Because the FCO does not have an event horizon the information loss problem is
resolved at the classical level. The elimination of black holes is the simplest
and possibly the least radical solution of the information loss puzzle.

It has been argued that the black hole information loss problem could be
resolved by an, as yet, unknown quantum gravity theory
\cite{bib:tHooft}.
However,
if it is argued that quantum gravity effects could remove
the black hole event horizon, then we are forced into a possibly unphysical
paradox, namely, an observer falling through a large black hole event
horizon with weak curvature would see entirely different physics associated
with the event horizon than an observer at a large distance from the
black hole. In order to overcome this paradox, long-range {\it non-local}
physics has to be postulated to exist, whereby an observer deep inside a
black hole can communicate to an outside observer the discrepancy detected
between their observations. As yet unproved claims have to be made
that what otherwise
appears to be a weak curvature event horizon for a large black hole,
which can be treated as a classical domain of gravity, is a
region of spacetime where quantum gravity effects associated with the
Planck mass describe the true physical situation. Moreover, all the information
inside the black hole must be encoded in the Hawking radiation
by an unknown dynamical mechanism.

The NGT that we have studied has been proved to
be perturbatively consistent for
weak gravitational fields
\cite{bib:Moffat_1,bib:Moffat_2}
It is asymptotically stable at future null
infinity
where the linear approximation is a stable solution of the field equations
without ghost poles or tachyons.
Higher-derivative theories can satisfy the regularity
theorems as applied to geodesics of the Riemannian geometry. This class of
theories can have repulsive gravity that retains the positive energy conditions
for physical matter
\cite{bib:Weyl,bib:Stelle,bib:Mannheim}.
There are no exact solutions of the field
equations in four dimensions analogous to the Wyman solution of NGT, so it is
not clear whether such theories can remove event horizons and black holes.
However, higher-derivative theories usually possess ghost poles, tachyons and
higher-order poles of one kind or another and this feature renders the
theories perturbatively unphysical. Theories with $R^2$
curvature terms in the Lagrangian exist, in which the metric tensor and the
connection are treated as independent variables
\cite{bib:Stelle,bib:Bahman}.
Ghost poles and
tachyons can be absent in these theories and they could be free of
singularities at $r=0$, but it is unlikely that they are free of black hole
event horizons.

The main result of this study is that classical
theories of gravity can be formulated that do not have singularities and
black holes. They satisfy the standard gravitational experimental tests.
Whereas it is feasible
that a quantum gravity program can be carried through successfully
in the future, we have demonstrated that a classical modification of EGT,
of the kind proposed here, can solve the singularity problem and the paradoxes
associated with black holes. We then avoid having to seriously modify quantum
mechanics or having to solve the difficult task of quantizing gravity
to resolve the black hole information loss problem.

The question remains to be answered: Does the non-singular, NGT
non-Riemannian geometry describe the true structure of spacetime, or is the
geometry of spacetime described by the Riemannian geometry associated
with NGT, with incomplete
geodesics and black holes? As we have seen, either description exists as a
possibility in NGT, and only future experiments can decide the outcome
of this fundamental question.
\vskip 0.2 true in
{\bf Acknowledgements}
\vskip 0.2 true in
I thank the Natural Sciences and Engineering Research
Council of Canada and the
Inter American Development Bank for their support of this research. I also
thank the Department of Physics, Cave Hill Campus, University of West Indies,
Barbados, for their hospitality while this work was being completed.
\vskip 0.3 true in

\end{document}